\documentclass[12pt]{article}  
\usepackage{a4wide,epsfig,amsmath,doublespace,array}
\newcommand{\beq}{\begin{equation}}
\newcommand{\eeq}{\end{equation}}
\newcommand{\bseq}{\begin{subequations}}
\newcommand{\eseq}{\end{subequations}}
\newcommand{\bear}{\begin{eqnarray}}
\newcommand{\eear}{\end{eqnarray}}
\newcommand{\preserveBackslash}[1]{\let\temp=\\#1\let\\=\temp}

\begin{document}

\title{Bifurcations and Chaos in Time Delayed Piecewise Linear Dynamical Systems}

\author{{\bfseries D.~V.~Senthilkumar and M.~Lakshmanan} \\
Centre for Nonlinear Dynamics,
Department of Physics, \\
Bharathidasan University,
Tiruchirappalli - 620 024, 
India}
\date{}


\maketitle
\begin{spacing}{2}

\begin{abstract} We reinvestigate the dynamical behavior of a first order
scalar nonlinear delay differential equation with piecewise linearity and
identify several interesting features in the nature of bifurcations and chaos
associated with it as a function of the delay time and external forcing
parameters.  In particular, we point out that the  fixed point solution
exhibits a stability island in the two parameter space of time delay and
strength of nonlinearity.  Significant role played by transients in attaining
steady state solutions is pointed out.  Various routes to chaos and  existence
of hyperchaos even for low values of time delay which is evidenced  by multiple
positive Lyapunov exponents are brought out.  The study is extended to the case
of two coupled systems, one with delay and the other one without delay.
\end{abstract}

\section{Introduction}
Chaotic phenomenon has been studied extensively in  several dynamical systems,
including Chua's circuits, cellular neural networks (CNNs) and cellular neural
networks with delay (DCNNs) during the past few decades.  Dynamical systems
described by delay difference-differential equations have many practical
applications in areas such as biology, secure communication, economics, traffic
control, neural networks, etc. For example,  Mackey and Glass[1977]  have
demonstrated  a model for blood production in patients with leukemia in terms
of a delay differential equation, where the concentration of blood can vary
chaotically when request for more blood is made.  Time delayed  dynamical
systems can also play a significant role in secure communications as these 
systems can have chaotic attractors with a large number of positive Lyapunov
exponents when the delay time of the  dynamical system is increased
[Goedgebuer et al., 1998] and so messages can be more securely transmitted through
chaotic synchronization.  Time delayed dynamical systems are essentially
infinite  dimensional systems, as an infinite set of independent numbers are
required to specify an initial condition.  Such infinite dimensional systems
can also be used to  forecast the fluctuation in the share market behavior
[Gwynne, 2001].  Various parameters in the traffic control theory are also
expressed in terms of delay coordinates [Helbing, 2001]. The dynamics of
delayed systems depends on the current state as well as on its past history and
recursively further past history and so on.  Thus CNN's with delay  coordinates
may be considered as ideal models for neural networks which act with their past
knowledge.

	It is a widely accepted fact that chaos can occur in autonomous time
continuous nonlinear systems  having order greater than two and in
nonautonomous time continuous nonlinear systems with order greater than one. In
time discrete systems chaos can occur even in first order invertible maps and
in non-invertible maps with order greater than one.  Recently Lu and He[1996]
have shown that chaos  can occur even in a simple scalar first order delayed
nonlinear (piecewise linear) dynamical system with large enough delay.  
Thangavel et al.[1998] have made preliminary study of the bifurcation scenario
and controlling of chaos in first order and coupled DCNNs  including the model
of Lu and He.

	In this paper, we have revisited the first order time delayed system
studied by Thangavel et al.[1998] with the addition of a constant external
forcing term and studied its dynamics in a large parameter regime of the 
external force and time delay. In particular, we have identified the novel
aspects of existence of a stable island in a two parameter space for the
equilibrium solutions  in the presence of time delay, the significant role of
transients in attaining steady state solution  and thereby identifying the
existence of familiar routes to chaos and the  emergence of hyperchaos even for small
values of time delay in a scalar first order piecewise linear delay
differential equation for a wide range of parameters, which is confirmed by
the existence of multiple positive Lyapunov exponents. The study has also been
extended to the case in which the first order scalar time delayed system is
coupled to a second scalar  system without delay.  Again the existence of a
stable island for equilibrium points is established.  Existence of different
bifurcation routes including type III intermittency route is pointed out, where
we have also discussed about the transient effects. We have also plotted the
two parameter bifurcation diagrams for both the cases to bringout the dynamics
in two parameter space. The plan of the paper is as follows. To start with,
a linear stability analysis is carried out both in the presence and in the absence of
time delay in sec.~2. In sec.~3, we have made a detailed study of  the effect
of transients in attaining the steady state solution leading to identification
of typical bifurcation scenario.  One and two parameter bifurcation diagrams
for different nonlinear parameter regimes characterized by the piecewise linear
function for different values of the parameter $p_2$ are studied in detail. We
also bringout the hyperchaotic nature of the systems with the existence of
multiple positive Lyapunov exponents. Further we have  considered the addition
of one more cell without delay in sec.~4, resulting in two cell DCNN and the
results of this system are discussed in sec.~5. The results are
summarized in sec.~6.
\section{Single cell Delayed Cellular Neural Network}
We consider the  following first order delay differential equation introduced
by Lu and He[1996], but with the  addition of a constant external force as
studied by Thangavel et al. [1998],
\bear
\frac{dx(t)}{dt}=-ax(t)+bf(x(t-\tau))+c, 
\eear
where $a$ and $b$ are parameters, $\tau$ is the time delay, $c$ is a constant external force and
$f$ is  an odd piecewise function defined as 
\bear
f(x(\tau))=
\left\{
\begin{array}{cc}
0,&  x\leq-p_2  \\
            -1.5x-2,&  -p_2 < x\leq -p_1 \\
            x,&    -p_1 < x\leq p_1 \\
            -1.5x+2,&  p_1 < x\leq p_2 \\
            0,&  x > p_2. \\ 
         \end{array} \right.
\eear
Here $p_1$ and $p_2$  are parameters. The form of the function $f(x(\tau))$ is sketched in Fig.~\ref{w1}.
In our study we have fixed the parameter $p_1$ at $p_1=0.8$
and studied the system behavior in a region of the external force $c\in[-0.15,0.15]$ and 
time delay $\tau\in[0.0,30.0]$  for different values of the parameter
$p_2$ characterizing the piecewise linear function $f(x(\tau))$. As any time delayed system 
depends strongly on its past history, which acts with its rich past 
knowledge, this model may be viewed as a simple one cell delayed cellular neural network (DCNN).
To start with let us consider the nature of the fixed (equilibrium) points of 
the system (1) in some detail.

\subsection{Fixed points and linear stability}
Equation (1) with the form (2)  for $f(x(\tau))$ can admit  upto a maximum of
three fixed  points, $x_i(t)=x_0,  i=1,2,3,$ depending upon the parameter
values. A linear stability  analysis of these fixed points under the linear
perturbation $x=x^*+\alpha\exp(\lambda t), \alpha \ll 1$, for Eq.~(1) was
carried out in [Thangavel et al., 1998]. However, the stability analysis in the
presence of time delay was done very briefly.  In this section,  we bring out
the existence of stable island in the $(\tau, 1.5b)$ plane for some of these
equilibrium points in the presence of  time delay in a detailed manner by
following the analysis of Ramana Reddy et al. [2000] for the
case of  coupled limit cycle oscillators with time delay coupling/time delay
feedback.  For this purpose we examine the stability nature of the fixed points
for Eq.~(1) both in the absence and in the presence of time delay $\tau$ by
considering the characteristic equation for the  eigenvalue $\lambda$ and
identifying the cases where Re$(\lambda)<0$ for stability. We consider two
cases.

\subsubsection{Time delay $\tau=0$}
 In the absence of time delay, the following fixed points can exist depending on
 the choice of parameters $a,b,c,p_1$ and $p_2$ in Eqs.~(1)-(2): \\
{\bf a)} For $|x| > p_2$, the fixed point is $x=x_0= \frac{c}{a}$  and the 
characteristic equation in this region is $\lambda=-a$.  The fixed point $x=x_0= \frac{c}{a}$
is stable for positive values of $a$.\\
{\bf b)} For $-p_2 < x \leq -p_1$, the fixed point is $x=x_0= \frac{c-2b}{a+1.5b}$  and the 
characteristic equation in this region becomes $\lambda=-(a+1.5b)$.  The fixed point 
is stable  when $a > -1.5b$.\\
{\bf c)} For $|x| \leq p_1$, the fixed point is $x=x_0=\frac{c}{a-b}$  and the characteristic 
equation becomes $\lambda=-(a+b)$.  The above fixed point is stable when  $a > -b$.\\
{\bf d)} For $p_1 < x \leq p_2$, the fixed point is $x=x_0= \frac{c+2b}{a+1.5b}$  and the 
characteristic equation in this region becomes $\lambda=-(a+1.5b)$.  The fixed point 
is stable  when $a > -1.5b$.

Next we consider the case when time delay is present, $\tau > 0$.
\subsubsection{Time delay $\tau>0$} 
{\bf a)}Here again, for $|x| > p_2$, the fixed point remains the same as for the case $\tau=0$, 
namely  $x=x_0= \frac{c}{a}$ and it is stable for $a > 0$.\\
{\bf b)} Next in the region, $-p_2 < x \leq -p_1$, for the fixed point $x=x_0=\frac{c-2b}{a+1.5b}$, the 
characteristic equation becomes the transcendental equation,
\beq
\lambda+a+1.5be^{-\lambda\tau}=0.  
\eeq
Let $\lambda=\alpha+i\beta$, where $\alpha$ and $\beta$ are real. Substituting
this into the  above  equation and equating real and imaginary parts, we obtain
equations for $\alpha$ and $\beta$ as
\beq
\alpha+a+1.5be^{-\alpha\tau}\cos(\beta\tau)=0, \\ 
\eeq
\beq 
\beta-1.5be^{-\alpha\tau}\sin(\beta\tau)=0.   
\eeq
From the above one can easily check by squaring and adding Eqs.~$(4)$ and $(5)$ that 
\beq
\beta=\beta_{\pm}=\pm\sqrt{2.25b^2\exp(2\alpha\tau)-(a+\alpha)^2}, \\
\eeq
and  
\beq
\alpha=-a-\frac{\beta}{\tan(\beta\tau)}.    
\eeq
Without loss of generality we choose + sign in the above equation, as the eigenvalues occur 
in complex conjugate pairs.  In order to 
find the critical stability curves, we choose $\alpha=0$.  Then we have
\beq
\beta|_{\alpha=0}=\pm\sqrt{2.25b^2-a^2}. \\
\eeq
From Eq.~(4), it follows that 
\beq
\beta\tau=\pm cos^{-1}\left(\frac{-a}{1.5b}\right)+2n\pi,   
\eeq
where n is any integer $(0,\pm1,\pm2,...)$.
Consequently one can expect that the stability regions are confined between the set of two curves
\begin{subequations}
\bear
\tau_1(n,1.5b)=\frac{2n\pi+\cos^{-1}(\frac{-a}{1.5b})}{\sqrt{2.25b^2-a^2}}, \\
\tau_2(n,1.5b)=\frac{2n\pi-\cos^{-1}(\frac{-a}{1.5b})}{\sqrt{2.25b^2-a^2}}. 
\eear
\end{subequations}
In Eq.~(10a), $n=0,1,2,...$ and in Eq.~(10b), $n=1,2,...$ for the pair of curves we have
chosen so that the curves have positive value of $\tau$.  To start with we note from
Eq.~(3) that when the time delay $\tau=0$, $\lambda=-a-1.5b$ and so $\alpha < 0$.
In order to identify those curves for $\tau > 0$ which
encompass the stable regions, the critical curves should be the ones on which 
$\frac{d\lambda}{d\tau} > 0$.  From Eq.~(3),
\beq
\frac{d\lambda}{d\tau} = \frac{ 1.5b\lambda\exp(-\lambda\tau)}{1-\tau1.5b\exp(-\lambda\tau)}
\eeq
and hence
\begin{eqnarray}
\frac{d\alpha}{d\tau}\bigg{|}_{\alpha=0} &=& Re\frac{1.5b(i\beta) \exp(-i\beta\tau)}
{1-\tau1.5b\exp(-i\beta\tau)}  \\  \nonumber
                            &=& \frac{1.5b\beta\sin(\beta\tau)}{D}   \\   \nonumber
			     &=& \beta^2D^{-1},
\end{eqnarray}
where
\begin{equation*}
D=\left[1-1.5b\tau\cos(\beta\tau)\right]^2 +
\left[1.5b\tau\cos(\beta\tau)\right]^2.
\end{equation*}
Therefore 
\begin{equation}
\frac{d\alpha}{d\tau} > 0         ~~\text{on both}~\tau_1~\text{and}~\tau_2.
\end{equation}
The above condition implies that there can be only one stability region between $\tau = 0$ line
(where $\alpha < 0$) and the critical curve $\tau_1(0,1.5b)$ which is the closest to the line 
$\tau = 0$. We note that the condition (13) prohibits the existence of any other stable region
(that is multistability regions) as in the case of time delayed limit cycle oscillators [Ramana
Reddy et al., 1998; 2000], because for a second stable region to exist one requires
$\frac{d\alpha}{d\tau} < 0$ on any one of the other curves $(n > 1)$. But this never occurs in
our case.  The numerical plot in Fig.~\ref{fig1a} of the curves  $\tau_1(n,1.5b)$ (solid curve for
$n=0,1,2$) and $\tau_2(n,1.5b)$ (broken curve for $n=1,2$) reveals that the region between $\tau =
0$ and  $\tau =\tau_1(0,1.5b)$ is  the only stable region (shaded region), where  $\frac{d\alpha}{d\tau}$ on
$\tau_1 > 0$, which passes from  negative to positive values of $\alpha$, whereas the other
curves $\tau_2(n,1.5b) < \tau < \tau_1(n,1.5b)$ for $n > 0$ do not satisfy the required
stability condition and hence they  are all associated with unstable regions.\\
{\bf c)} As in the previous region, for the case  $-p_1 < x \leq p_1$ and for the fixed point
$x_0=\frac{c}{a-b}$, the characteristic equation in this region has the form
\beq
\lambda+a+be^{-\lambda\tau}=0.   
\eeq
As before, we obtain a set of critical curves  as 
\bseq
\bear
\tau_1(n,b)=\frac{2n\pi+\cos^{-1}(\frac{a}{b})}{\sqrt{b^2-a^2}},  \\
\tau_2(n,b)=\frac{2n\pi-\cos^{-1}(\frac{a}{b})}{\sqrt{b^2-a^2}},  
\eear
\eseq
where $n=0,+1,+2,...$ in Eq.~(15a) and $n=+1,+2,...$ in Eq.~(15b). Also we get 
\begin{equation*}
\frac{d\alpha}{d\tau}\bigg{|}_{\alpha=0} =  \beta^2D^{-1}
\end{equation*}
where
\begin{eqnarray*}
D &=&\left[1-1.5b\tau\cos(\beta\tau)\right]^2 + \left[1.5b\tau\cos(\beta\tau)\right]^2\\
\beta &=& b\sin(\beta\tau)
\end{eqnarray*}
with the same argument as for the above case, we find that there is only one stable island 
between the curves $\tau = 0$ and $\tau_1(0,b)$.\\
{\bf d)}For the case $p_1 < x \leq p_2$ and for the fixed point $x=x_0= \frac{c+2b}{a+1.5b}$,
the characteristic equation turns out to be identical to Eq.~(3) and the associated
critical curves also have similar form as that of Eq.~(10). 
\section{Numerical study of the single cell DCNN}
In this section, we will present a discussion of the dynamics of the delayed
cellular neural network (Eq.~(1)) in the pseudospace $(x(t),x(t-\tau))$.
Further we  will also discuss the significant role played by transients in
attaining the steady state behavior, the computational efficiency required for
achieving  such steady state solution and the nature of bifurcation diagrams
for both low and high transients with the external forcing $c$ as the control
parameter.  We will also point out the existence of a \emph{stable} island for
equilibrium points in the two parameter (now in the time delay $\tau$ - the
external forcing $c$ plane) bifurcation diagrams for various ranges of control
parameters and nonlinearity characterized by the function $f(x(\tau))$ for
three different values of $p_2$ in Eq.~(2). In addition, we have also
calculated the Lyapunov exponents associated with the system and show that
there exists a large parameter range over which hyperchaos exists
(corresponding to multiple positive Lyapunov exponents). From the two parameter
bifurcation diagrams, it becomes evident that the sizes of the hyperchaotic
regime and the stable fixed point regime increase with time delay.
\subsection{Dynamics in the pseudospace}
The dynamics of the DCNN can be studied in a suitable phase space by plotting
the numerical solution of Eq.~(1)  appropriately.  Now, to calculate $x(t)$
from  Eq.~(1) for times greater than $t$, the  function $x(t)$ over  the
interval $(t-\tau,t)$ must be given.  Hence, for a prescribed continuous
function $x(t)$ on $(-\tau,0)$ one can integrate Eq.~(1) using numerical
methods as in the case of ordinary differential  equations.  We have
numerically integrated Eq.~(1) using Runge-Kutta fourth order integration
routine with the parameters $a=1.0,b=1.2$ and $p_1=0.8$  for three different
values of $p_2=1.0,1.33,1.66$ with the initial  condition $x(t)=0.9$ in the
interval $(-\tau,0)$. As the system is of first order in nature, the dynamics
can be viewed in a pseudospace by plotting $x(t)$ against  $x(t-\tau)$.  (Note
that  the choice of $x(t-\tau)$ as the second phase variable is arbitrary;
$x(t-t^\prime)$ can be equivalently used, where $t^\prime$ is an arbitrary time
delay).  One encounters typical scenario of  bifurcations leading to chaos, but
with an important difference: \emph{Transients play a crucial role} and it
takes a very long time before transients die down to attain steady state
solutions. In particular, period doubling and inverse period doubling
phenomena, besides other bifurcations, are often encountered. In many cases the
system exhibits  \emph{hyperchaotic behavior} characterized by multiple
positive Lyapunov exponents, which we have discussed below in sec.~3.4.
Fig.~\ref{fig2} shows a typical  pseudochaotic attractor for $\tau=5.0,c=0.001$
and $p_2=1.33$ with the initial condition as given above in the phase  space of
$x(t)$ against $x(t-5.0)$. The positive maximal Lyapunov exponent of this
attractor for the above mentioned parameter values is $\lambda_{max}=0.05461$.
It may be noted that previously it was reported by Thangavel et al.~[1998] and Lu
and He~[1996] that the system  for a different set of parameters exhibits only
chaos  (and not hyperchaos) for large values of time delays , that is $\tau >
20$.  On the otherhand we have identified the parameter regimes where the
system exhibits chaos and even hyperchaos for small values of time delay around
$\tau = 5.0$.  In the following we present the details.

\subsection{Transients}
Usually dynamical systems described by ordinary differential equations can attain the steady
state  within a few thousand transients for optimal value of time step $\Delta t$, provided  due
care is given for numerical accuracy and suitable numerical algorithm is chosen.  However for the
delay differential equation  considered in this paper, such  low transients (of the order $10^4$
and more) do not seem to lead the system to their steady state solutions. In such cases one has
to leave out more number of transients so that one can have the possibility of obtaining a clear
picture of the bifurcation diagram with the existing routes.  In fact, the effect of transients
has been pointed out by Becker and Dorfler[1989] even for the logistic map. As for as nonlinear
maps and ordinary differential equations are concerned, with some effort it is possible to
realize whether the system has reached the steady state or not by constructing the appropriate
bifurcation diagrams and identifying the various bifurcation routes  by leaving sufficient number
of transients before starting the analysis for steady state solutions.  The same considerations
hold good for dynamical systems modeled by  finite number of coupled nonlinear maps and odes
without time delay.

On the other hand, when one starts studying nonlinear dynamical systems with time delay,
effectively one considers infinite number of coupled odes which in a numerical sense corresponds
to several hundred coupled nonlinear odes.  For example, for Eq.~(1), when the time delay
$\tau=25.0$,  in the  actual numerical analysis, we  typically take the optimal time step as
$\Delta t=0.05$, so that we are actually solving 500 coupled nonlinear odes. However, the optimal
time step $\Delta t$ to be fixed depends explicitly on the nature of the system and on the time
delay introduced in the system.  In contrast to the nonlinear odes, where either a very small  or
a relatively large time step $\Delta t$ results in an exponentially increasing  numerical error,
for the time delayed system (1), we have found that the optimal time step $\Delta t$ varies 
between $0.05$ to $0.0001$, depending on which the actual number of coupled nonlinear odes
increase, and the number of transients to be left out is found to  lie anywhere between $2.0-3.0
\times 10^6$. For small values of time delay $\tau$, the optimal time step $\Delta t$ should be
as small as possible to identify a typical bifurcation scenario.  However, for large values of
time delay $\tau$,  a very  small value of time step $\Delta t$ can again be problematic as one
has to deal with a large number of coupled odes and the time required for
solving such huge number of coupled odes becomes huge. Also a   large value of time step $\Delta t$ does not
seem to lead to the steady state solution,   as such time steps lead to propagation of numerical
errors.  However, an appropriate optimal choice of time step $\Delta t$ can  lead to a typical
bifurcation scenario.

Thus the numerical analysis of time delayed nonlinear systems requires an appropriate time step
$\Delta t$ to be chosen, depending on which the actual number of coupled nonlinear odes
increases, which in turn necessarily increases the number of transients to be left out in order
to obtain the steady state solution.  In such cases computing with a standard personal computer
like a 32bit Pentium IV processor is a difficult and time consuming task.   However, it turns out
that with a high speed computers like an Alpha workstation, such a task becomes feasible. We have
carried out our analysis with a 64 bit Alpha workstation, in which it took 50-60 days to plot
each of the two parameter bifurcation diagrams leaving out sufficient transients using Runge-Kutta
fourth order integration routine.


In our numerical analysis we have chosen the optimal time step as $\Delta
t=0.05$ for the value of time delay $\tau=25.0$.  The effect of transients in
reaching the steady state behavior can be realized from the bifurcation
diagrams we have obtained for the parameter values $a=0.16, b=0.2$ and for two
different values of $p_2$. Fig.~\ref{fig2a}a shows the one parameter
bifurcation diagram for the above mentioned parameter values with $p_2=1.33$,
when we leave  transients of the order $1.0 - 2.0 \times 10^4$, which is
sufficiently large in the case of nonlinear maps and odes, leaving the
impression that the steady state has been reached with a bifurcation scenario
that is quite complicated and atypical. In contrast, Fig.~\ref{fig2a}b shows
the bifurcation diagram for the same value of the parameters except that now
the transients left out are of the order of $1.0 \times 10^5$, which shows a
typical bifurcation scenario. Similarly Fig.~\ref{fig3}a shows the bifurcation
diagram for the same values of parameters as above for $\tau, a$ and $b$ with 
$p_2=1.66$ for the transients of the order $1.0 \times 10^5$, which shows a
very complex structure, whereas Fig.~\ref{fig3}b shows the bifurcation diagram
for the transients of the order $1.4 \times 10^6$, in which case still there
exists some stray points near the bifurcation regions, which indicates that the
system requires still more transients to settle to its steady state.  The
complexity increases with time delay and hence the number of transients  also
increases, which in turn increases the computing time enormously. This is also
evident from Fig.~\ref{lypbif}, where the maximum value of $x(t)$, XMAX, is
plotted against the time delay $\tau$, for the parameter values  $a=1.0, b=1.2,
c=0.001$ and $p_2=1.33$.  Fig.~\ref{lypbif}a shows that even for transients of
the order $1.0\times 10^5$ , the bifurcation diagram is still in an unsettled
form for larger values of time delay $\tau$.  On the other hand in
Fig.~\ref{lypbif}b, the number of transients is of the order $2.5\times 10^5$,
wherein a clear bifurcation scenario has emerged. We have verified the role of
transients to attain  the steady state solutions for various values of time
step $\Delta t$ and time delay $\tau$ for few other delay differential
equations also, including those discussed in the work of  Ramana Reddy et al.,
[2002].  The results of  our detailed numerical investigation for Eq.~(1) are
tabulated in Table I, where we have indicated the number of transients to be
left out for given values of $\Delta t$ in order to attain the steady state
solutions.

\begin{center}
Table I: Effect of step-size on transients
\begin{tabular}{|c|c|c|c|}
\hline 
Value of $\tau$ & 
Optimal value of $\Delta t$ & 
Number of coupled & 
Number of transients\\
& &  differential equations & to be left out \\\hline
$5$ & $0.002$ & $250$ & $3.0-5.0 \times 10^5$ \\
    & $0.001$ & $500$ & $5.0-6.0 \times 10^5$ \\
$10$ & $0.002$ & $5000$ & $1.0-2.0 \times 10^6$ \\
    & $0.001$ & $10000$ & $ > 2.0 \times 10^6$ \\
$25$ & $0.05$ & $500$ & $1.0 \times 10^5-2.0 \times 10^6$ \\
     & $0.005$ & $5000$ & $1.0-2.0 \times 10^6$ \\
     & $0.0025$ & $10000$ & $ > 2.0 \times 10^6$ \\ \hline
\end{tabular} 
\end{center}

As the time delay increases, the size of the delay loop (that is to be updated
on every iteration) required to maintain the delay variable in  numerical
simulation also increases, which in turn increases the computational effort. 
Another important point to note in connection with the effect of transients is
the following.  In the above we have discussed the effect with reference to a
scalar delay differential equation only.  On the other hand, in recent times,
there has been considerable interest to study the dynamics of multiply coupled
delayed neural  networks in biological systems [Campbell et al., 1999; Earl \&
Strogatz, 2003]. In order  to examine the actual steady state behavior of such
networks, which  effectively corresponds to several thousands of coupled
nonlinear odes, the time  required to obtain steady state solution increases
enormously as each one of them have  separate time delay and intrinsic
transient behavior and it becomes crucial to study  the effect of the later.

\subsection{One and two parameter bifurcation diagrams}  
All the bifurcation diagrams shown in this paper are plotted after leaving out
a very large  number transients of of the order of $1.2-3.2 \times 10^6$.
Fig.~\ref{fig2a}b shows the one parameter $(c)$ bifurcation diagram 
characterized by the piecewise linear function $f(x)$  in Eq.~(2) with
$p_1=0.8, p_2=4/3,\tau=25.0$ for  $c\in[-0.1,-0.05]$. It clearly exhibits
period doubling scenario; however there exists a sudden distortion around the
value of $c=-0.0785$, the cause of which remains unexplained by any standard
type of bifurcation. (We wish to point out that the  numerical analysis in
[Thangavel et al., 1998] actually corresponds to the value of $p_2=1.0$ and  not
$p_2=4/3$ as stated in that paper). Similarly Fig.~\ref{fig3}b shows the 
bifurcation diagram with the values $p_1=0.8, p_2=5/3, \tau=25.0$ within the
range of   $c\in[-0.15,-0.136]$, exhibiting reverse period doubling route to
chaos in the range of $c\in[-0.136,-0.144]$ that includes a  clear band merging crises and anti-monotonicity in the
range of  $c\in[-0.15,-0.144]$.

The two parameter bifurcation diagram for $\tau\in[0,30]$ and
$c\in[-0.16,0.16]$ when $p_2=1.0$  is shown in Fig.~\ref{fig4}a, which shows
the behavior of the DCNN for the combined phase  space of parameters $\tau$ and
$c$.  The following colour codes are used to represent various  regions,
period-1 region -red, period-2 region - green, 3-blue, 4-yellow, 5-magenta,
6-cyan,  7-gray, 8-copper, chaos-black and the fixed points-white.  The white
region in the two parameter  bifurcation diagram corresponds to stable regions.
Fig.~\ref{fig4}b shows the two parameter bifurcation diagram for the same
parameters as in Fig.~\ref{fig4}a, except  that now $p_2=4/3$.  We use the same
colour codes as in Fig.~\ref{fig4}a for all the two  parameter bifurcation
diagrams in this paper.  Similarly Fig.~\ref{fig4}c shows the two  parameter
bifurcation diagram for the same range of the control parameters and the  same
values of the parameters $a$ and $b$ in Eq.~(1) except that now $p_2=5/3$. By
comparing the Figs.~\ref{fig4}a,~\ref{fig4}b  and~\ref{fig4}c, we can see that
the  stable fixed point regions and the chaotic regions  increase for  small
changes in the nonlinear parameters  $p_1$ and $p_2$.  Further we  infer from 
these figures that as the delay increases the chaotic nature of the system also
increases and thereby contributing to the  hyperchaotic  nature of the system
characterized by  multiple positive Lyapunov exponents. 
\subsection{Lyapunov exponents and hyperchaotic regimes}
One of the interesting aspects of the dynamics associated with Eq.~(1) is the
existence of hyperchaos in a single first order scalar equation with time delay
even for small values of the latter  for suitable values of other system
parameters, whereas mention was made in the earlier papers
[Lu~\&~He,~1996;~Thangavel et al.,~1998] that the system Eq.~(1) exhibits only
simple  chaos even for large values of time delay.  As Eq.~(1) is  a first
order delay  differential equation, the usual procedure for calculating
Lyapunov exponents is not applicable. However, the simple idea of approximating
a single scalar differential equation with delay by an N-dimensional discrete
mapping [Farmer;~1982], which is of infinite dimensions, facilitates one to
calculate the Lyapunov exponents.  To stimulate the behavior  of such systems
on a computer it is necessary to approximate the continuous evolution of an 
infinite dimensional system by a finite number of elements whose values change
at discrete  time steps.  In this manner a continuous infinite dimensional
dynamical  system is replaced by a finite dimensional iterated mapping.  This
method was proposed  originally by Farmer[1982] to calculate the Lyapunov
exponents for delay systems.  As the  delay parameter is increased, for most
parameter values the dimension increases and the  attractor generally becomes
more complicated, thereby contributing to  the hyperchaotic nature of the
system, which gets confirmed by the increasing  number  of positive Lyapunov
exponents. The first ten maximal  Lyapunov exponents, for the parameter values
$a=1.0,b=1.2, c=0.001, p_1=0.8, p_2=1.33$  and $\tau \in(2,29)$ is shown in
Fig.~\ref{fig1b}, where it is evident that the number  of positive Lyapunov
exponents increases with time delay $\tau$. Almost the entire black regime in
the two parameter bifurcation diagrams Fig.~\ref{fig4}  for large values of
$\tau$ is characterized by multiple positive Lyapunov exponents  contributing
to hyperchaotic nature of the system.
\section {Stability analysis and chaotic dynamics of two cell DCNN}
 We now add one new cell without delay to the Eq.~(1) to obtain the following coupled equation
\bseq
\bear
\frac{dx(t)}{dt}&=&-ax(t)+bf(x(t-\tau))+dy(t),  \\
\frac{dy(t)}{dt}&=&-cy(t)+ex(t),               
\eear
\eseq
where $c$ and $e$  are additional parameters and the function $f(x)$ and other
parameters are the same as already  defined  in Eq.~(1).   As in the case of
single cell DCNN, we have fixed the parameter $p_1$  at $p_1=0.8$ and studied
the system behavior in a range of variable parameter $c\in[0.0,1.4]$ and time
delay $\tau\in[0.0,30.0]$ for different values of the parameter $p_2$
characterizing the piecewise linear function $f(x)$.  Eq.~(16) can also be
viewed as a two cell DCNN, one cell  acting with its past knowledge coupled
with another cell which does not have any past knowledge.  Now let us consider
the nature of the fixed points of system (16).
\subsection{Fixed points and Linear stability}
We will bring out the existence of stable island in two cell DCNN by linear
stability  analysis in the presence of time delay $\tau$, as in the case of
single cell DCNN.   For this purpose, we examine the stability nature of the
fixed points of Eq.~(16)  both in the absence and in the presence of time delay
$\tau$.
\subsubsection{Time delay $\tau=0$}
In the absence of time delay, the following fixed points can exist depending 
on the choice of parameters $a,b,c,d,e,p_1$ and $p_2$ in Eqs.~(16) and (2).\\
{\bf a)} For $|x| > p_2$, the fixed point is $(x,y)=(x_0,y_0)=(0,0)$  and the 
characteristic equation in this region is $\lambda^2+(a+c)\lambda+ac-de=0$.  
The fixed point $(x,y)=(0,0)$ is stable when $a+c>0$ and $ac>de$.\\
{\bf b)} For $-p_2 < x \leq -p_1$, the fixed point is $(x,y)=(\frac{2bc}{de-(a+1.5b)c},
\frac{2be}{de-(a+1.5b)c})$   and the 
characteristic equation in this region becomes $\lambda^2+(c+a+1.5b)\lambda+(ca+1.5bc-de)=0$.
The fixed point is stable  when $c+a+1.5b>0$ and $ac+1.5bc>de$.\\  
{\bf c)} For $|x| \leq p_1$, the fixed point is $(x,y)=(0,0)$   and the characteristic 
equation becomes $\lambda^2+(c+a-b)\lambda+(ca-bc-de)=0$.  The above fixed point is 
stable when  $c+a-b>0$ and $ac>bc+de$.\\
{\bf d)} For $p_1 < x \leq p_2$, the fixed point is  $(x,y)=(\frac{-2bc}{de-(a+1.5b)c},
\frac{-2be}{de-(a+1.5b)c})$    and the 
characteristic equation in this region becomes $\lambda^2+(c+a+1.5b)\lambda+(ca+1.5bc-de)=0$.  
The fixed point is stable  when $c+a+1.5b>0$ and $ac+1.5bc>de$.\\  
Next we consider the case when time delay is present, $\tau > 0$.
\subsubsection{Time delay $\tau>0$} 
{\bf a)} For $|x| > p_2$, the fixed point  and its stability remains the same as for the case 
$\tau=0.0$\\
{\bf b)} Next in the region $-p_2 < x \leq -p_1$, for the fixed point $(x,y)=(\frac{2bc}
{de-(a+1.5b)c},\frac{2be}{de-(a+1.5b)c})$  the characteristic equation is
\begin{equation}
(c+\lambda)(\lambda+a+1.5b\exp(-\lambda\tau))- de = 0,
\end{equation}
which is a transcendental equation with infinite number of solutions. Let $\lambda=\alpha+i\beta$,
where $\alpha$ and $\beta$ are real.  Substituting this in the above equation and equating real
and imaginary parts, we obtain two equations as
\bseq
\bear
(c+\alpha)(\alpha+a+1.5b\exp(-\alpha\tau)\cos(\beta\tau))-de-\beta^2+1.5\beta b\exp(-\alpha\tau)\sin(\beta\tau)=0, \\
(c+\alpha)(\beta-1.5b\exp(-\alpha\tau)\sin(\beta\tau))+\beta(\alpha+a+1.5b\exp(-\alpha\tau)\cos(\beta\tau))=0
\eear
\eseq
In order to find the critical stability curves, we choose $\alpha=0$.  Then we have
\bseq
\bear
\label{eqonea}
1.5bc\cos(\beta\tau)+ca-de-\beta^2+1.5\beta b\sin(\beta\tau)=0, \\
1.5b\beta\cos(\beta\tau)+c\beta+\beta a -1.5bc\sin(\beta\tau)=0.
\label{eqone}
\eear
\eseq
Multiplying the above two equations (\ref{eqonea}) and (\ref{eqone}) with $c$ and $\beta$,
respectively, adding we obtain
\beq
1.5b\cos(\beta\tau)(c^2+\beta^2)+a(c^2+\beta^2)-cde=0.
\label{eqtwoa}
\eeq 
Now multiplying the equations (\ref{eqonea}) and (\ref{eqone}) with $\beta$ and $c$, 
respectively, subtracting the resulting equations, we obtain
\beq
1.5b\sin(\beta\tau)(c^2+\beta^2)-\beta(c^2+\beta^2)-\beta de=0.
\label{eqtwob}
\eeq
Squaring and adding the equations (\ref{eqtwoa}) and (\ref{eqtwob}), rearranging 
them, we obtain the following cubic equation for $\beta^2$,
\beq
X^3+uX^2+vX+w=0, X=\beta^2,
\label{eqthr}
\eeq
where the constants are given as
\begin{equation*}
u=a^2+2(de+c^2)-2.25b^2;
v=(de+c^2)^2-2ac(de-ac)-4.5b^2c^2;
w=c^2(de-ac)^2-2.25b^2c^4.
\end{equation*}
From Eq.~(\ref{eqtwoa}), we obtain
\beq
\beta\tau=
\pm\cos^{-1}\left(\frac{cde-a(c^2+\beta^2)}{1.5b(c^2+\beta^2)}\right)+2n\pi,
\eeq
where $n$ is any integer $(0,\pm1,\pm2,...)$ and the value of $\beta$ can be
obtained by solving the Eq.~(\ref{eqthr}) for $\beta^2$. Consequently the stability regions 
are confined between the set of curves,
\bseq
\bear
\label{eqfoura}
\tau_1(n,1.5b)=\frac{2n\pi+\cos^{-1}\left(\frac{cde-a(c^2+\beta^2)}{1.5b(c^2+\beta^2)}\right)}{\beta}  \\
\tau_2(n,1.5b)=\frac{2n\pi-\cos^{-1}\left(\frac{cde-a(c^2+\beta^2)}{1.5b(c^2+\beta^2)}\right)}{\beta},
\label{eqfourb}
\eear
\eseq
where $n=0,+1,+2,...$ in Eq.~(\ref{eqfoura}) and $n=+1,+2,...$ in
Eq.~(\ref{eqfourb}).  In order to
check whether the region enclosed by the curve $\tau_1$ and $\tau_2$ for every
value of $n$ forms stable islands in $(\tau,\beta)$ plane, one has to
examine the sign of $\frac{d\alpha}{d\tau}$  on $\tau_1$ and $\tau_2$ as we
have done for the case of single cell DCNN. We have found that  there exists
only one stable island between  $\tau=0$ and $\tau=\tau_1(0,\beta)$ curve
(shaded region) as shown in Fig.~\ref{figad}.\\
{\bf c)} In the region $|x|\leq p_1$, for the fixed point $(x,y)=(0,0)$, the characteristic equation is
\begin{equation*}
(c+\lambda)(\lambda+a+b\exp(-\lambda\tau))- de = 0,      
\end{equation*}
is a transcendental equation with infinite number of solutions. Proceeding in the same way as for the
case $-p_2 < x \leq -p_1$, we obtain a set of critical curves as
\bseq
\bear
\label{eqfivea}
\tau_1(n,b)=\frac{2n\pi+\cos^{-1}\left(\frac{cde-a(c^2+\beta^2)}{b(c^2+\beta^2)}\right)}{\beta}  \\
\tau_2(n,b)=\frac{2n\pi-\cos^{-1}\left(\frac{cde-a(c^2+\beta^2)}{b(c^2+\beta^2)}\right)}{\beta},
\label{eqfiveb}
\eear
\eseq
where $n=0,+1,+2,...$ in Eq.~(\ref{eqfivea}) and $n=+1,+2,...$ in
Eq.~(\ref{eqfiveb}).  As for the previous case
we found that there is only one stable regions between $\tau=0$ and $\tau=\tau_1(0,\beta)$. \\
{\bf d)}For the case $p_1 < x \leq p_2$ and for the fixed point $(x,y)=(\frac{-2bc}{de-(a+1.5b)c},
\frac{-2be}{de-(a+1.5b)c})$ ,
the characteristic equation turns out to be indentical to Eq.~(17) and the
associated critical curves also have similar form as that of
Eqs.~(\ref{eqfoura}) and ~(\ref{eqfourb}). 
\section{Numerical study the two cell DCNN}
In this section, we will discuss the dynamics of the two cell DCNN defined by
Eq.~(16) through  numerical  analysis and the transient effects involved in it
by studying the bifurcation diagrams  for both low and high transients with $c$
as control parameter.  The one parameter  bifurcation diagram in one of the
regimes has the signature of type III  intermittent  behavior. We have also
found that the dynamics at this transition possesses type III  intermittent
characteristic scaling behavior. We have also plotted the two parameter
bifurcation diagrams  in the $(\tau,c)$ plane for three different values of
$p_2$, from which  also we have pointed out the  existence of stable island for
equilibrium points for suitable choice of other parameter values.
\subsection{Transients}
As we have already discussed much about the role of transients in the case of
single cell DCNN, in Sec.~3.2, it is not necessary to discuss it once again in
detail for the present case. So we point out only the number of transients
required to obtain the bifurcation  diagrams which we have shown in
Figs.~\ref{figccwdb1} and ~\ref{figccwdb2} for the system(16), which enables
one to realize the effect of transients.  As discussed in the case of single
cell DCNN, the transients predominates evolution of the system(16) to its
steady state solutions.   Fig.~\ref{figccwdb1}a shows the one parameteriii
bifurcation diagram for transients of the order  $2.0 \times 10^5$ for the
parameter value $p_2=1.33$, whereas Fig.~\ref{figccwdb1}b shows  the one
parameter bifurcation diagram for transients of order $1.75 \times 10^6$ for
the  same parameter values. The stray points in the neighbourhood of
bifurcation points are due  to the transient effects, which suggests that it
requires still more iterations need to be left out.   Fig.~\ref{figccwdb2}a
shows the one parameter bifurcation diagram  for transients of the order $2.0
\times 10^5$ for the parameter value $p_2=1.33$, whereas  Fig.~\ref{figccwdb2}b
shows the  bifurcation diagram for transients  of the order $2.0 \times 10^6$ 
for the fixed values of other parameters.
\subsection{One and two parameter bifurcation diagrams}
We have integrated Eq.~(16) with the parameters $a=0.16, b=0.2, d=0.2, e=0.2$
and $c$ as variable parameter with the initial conditions  $x(t) = 0.9$ for
$t\in(-25,0)$ and $y=0.8$. Fig.~\ref{figccwdb1}b shows the one parameter 
bifurcation diagram  in the nonlinear parameter regime characterized by the
function  $f(x(\tau))$ with $p_1=0.8,p_2 = 4/3$ and $\tau=25.0$, which shows 
the period-3 doubling bifurcation route to chaos.   Fig.~\ref{figccwdb2}b shows
reverse  period doubling in the range $c\in[0.32,0.345]$, and at the critical
value of  $c_{crit}=0.345969$ the system exhibits intermittent transition to
chaos followed by reverse period 5 doubling in the range $c\in[0.346,0.365]$
interspersed  by periodic windows for the same parameter values as above except
that now $p_2=5/3$. At the intermittent transition, the amplitude variation
loses its regularity and  a burst appears in the regular phase as shown in
Fig.~\ref{figint}.  This behavior repeats  as time increases as observed in the
usual type III intermittent scenario.  The duration of laminar phases is random
during the transition and finally results  in chaotic oscillations by
increasing the critical value of the control parameter $c$.  The plot of mean
laminar length $<l>$ as a function of the parameter $f=(c_{crit}-c)$ is shown
in Fig.~\ref{figpl}, where $c_{crit}$ is the critical value of the parameter
for the occurrence of the intermittent transition.   The phase space
trajectories reveal a power law relationship of the form $<l>=f^{-\alpha}$ with
the estimated value of $\alpha=0.871$.  This analysis confirms that the
trajectories at the critical value of $c$ is  associated with standard
intermittent dynamics of type III described in the work of Pomeau \& Manneville
[1980] and Schuster [1988]. Figs.~\ref{fig6} shows the global bifurcation
diagrams when  $p_2 = 1.0, 4/3 $ and $5/3$ respectively.  The colour codes used
here are the same as in the previous  case of single cell DCNN.  By comparing
the global bifurcation diagrams, we can realize that the stable island (fixed
point) and the  chaotic region increases for small changes (increase) in the
parameter $p_2$.  In addition the chaotic  nature of the system increases with
delay.

\section{Conclusion}
We have studied in detail the linear stability nature of equilibrium points,
both in the absence and in the presence of  time delay, of the DCNN systems and
identified the existence of stable island in a single scalar piecewise linear
differential equation with inherent time delay.  The significant role of
transients  in attaining the steady state behavior  has also been brought out. 
One and two parameter bifurcation diagrams for specific choice of parameters
have been presented in support of our results.   Hyperchaotic nature of the
system, even for small time delay in a simple first order piecewise linear
differential equation ensures that such systems can be used for highly
efficient secure communication by constructing equivalent electronic circuits. 
Therefore based on our results the respective circuits with suitable delay can
be designed and can be tested for specific applications.  These studies have
also been extended to coupled systems with time delay and the generic nature of
the results established.

\section*{Acknowledgments}
This work forms part of a Department of Science and Technology, Government of
India sponsored research project.
\end{spacing}
\newpage
\section*{References}
\begin{enumerate}

\item
Becker,~K.~H.~\&~Dorfler,~M. [1989]~{\it{Dynamical systems and fractals}}~ 
(Cambridge University Press, Cambridge).

\item
Campbell,~S.~A.,~Ruan,~S.~\&~Wei,~J.~[1999]~``Qualitative Analysis of a 
Neural Network Model with MultipleTime Delays," {\it{Int. J. of Bifurcation and chaos}} 
{\bf 9}(8), 1585-1595.

\item
Earl,~M.~G.~\&~Strogatz,~S.~H.~[2003]~``Synchronization in Oscillator Networks
with Delayed Coupling:A Stability Criterion," {\it{Phys. Rev. E.}} {\bf 67}, 036204.

\item
Farmer,~J.~D.~[1982]~``Chaotic Attractor of an Infinite-Dimensional Dynamical System,"
{\it{Physica 4D}}, 366-393.

\item
Goedgebuer,~J.~P.,~Larger,~L.~\&~Porte,~H.~[1998]~``Optical Cryptosystem Based on 
Synchronization of Hyperchaos Generated by a Delayed Feedback Tunable Laser Diode,"
{\it{Phys. Rev. Lett.}} {\bf 80},  2249-2252.

\item
Gwynne,~P.~[2001]~``Physicist who makes Cash from Chaos," {\it{Physics World}} {\bf 9}, 9.

\item
Helbing,~D.~[2001]~``Traffic and Related Self-driven Many-particle Systems," 
{\it{Rev. of Mod. Phys.}} {\bf 73}, 1067-1141. 

\item
Herrero,~R.,~Figueras,~M.,~Rius,~J.,~Pi,~F.,~\&~Orriols,~G.~[2000]~``Experimental
Observation of the Amplitude Death Effect in Two Coupled Nonlinear Oscillator,"
{\it{Phys. Rev. Lett.}} {\bf 84}, 5312-5316.

\item
Ikeda,~K.~\&~Matsumato,~K.~[1987]~``High-dimensional Chaotic Behavior in Systems
with Time Delayed Feedback,"{\it{Physica D}} {\bf 29},  223-235.

\item
Kim,~S.,~Park,~S.~H.,~\&~Ryu,~C.~S.~[1997]~``Multistability in Coupled Oscillator
Systems with Time Delay,"{\it{Phys. Rev. Lett.}} {\bf 79}, 2911-2915.

\item
Lim,~T.~K.,~Kwak,~K.~\&~Yun,~M.~[1998]~``An Experimental Study of Storing Information
in a Controlled Chaotic System with Time Delayed Feedback,"{\it{Phys. Lett. A}} {\bf 240}, 
287-294.

\item
Lu,~H.~\&~He,~Z.~[1996]~``Chaotic Behavior in first Order Autonomous Continuous
Time System with Delay,"{\it{IEEE Trans. Circuits Syst.I}} {\bf 43}, 700-702.

\item
Mackey,~M.~C.~\&~Glass,~L.~[1977]~``Oscillation and Chaos in Physiological Control Systems,"
{\it{Science}} {\bf 197}, 287-289.

\item
Ott,~E.~[1993]~{\it{Chaos in dynamical systems}}~(Cambridge University Press, Cambridge).

\item
Pomeau,~Y.~\&~Manneville,~P.~[1980]~``Intermittent Transition to Turbulence in
Dissipative System,"{\it{Commun. Maths. Phys.}} {\bf 74}, 189-197.

\item
Ramana Reddy,~D.~V.,~Sen,~A.~\&~Johnston,~G.~L.~[1998]~``Time Delay Induced Death in 
Coupled Limit Cycle Oscillators,"{\it{Phys. Rev. Lett.}} {\bf 80}, 5109-5113.

\item
Ramana Reddy,~D.~V.,~Sen,~A.~\&~Johnston,~G.~L.~[2000]~``Experimental Evidence of
Time Delay Induced Death in Coupled Limit Cycle Oscillators,"{\it{Phys. Rev. Lett.}} 
{\bf 85}, 3381-3385.

\item
Schuster,~H.~G.~[1988] {\it{Deterministic Chaos}}~(VCH Publishers, Weinheim).

\item
Thangavel,~P.,~Murali,~K.~\&~Lakshmanan,~M.~[1998]~``Bifurcation and Controlling of 
Chaotic Delayed Cellular Neural Networks,"{\it{Int. J. of Bifurcation and chaos}} {\bf 8},
2481-2492.

\item
Yaowen,~L.,~Guangming,~G.,~Hong,~Z.~\&~Yinghai,~W.~[2000]~``Synchronization
of Hyperchaotic Harmonics in Time Delay Systems and its Application to Secure
Communication,"{\it{Phys. Rev. E.}} {\bf 62}, 7898-7904.

\end{enumerate}

\newpage
\begin{figure}[!ht]
\caption{The form of the piecewise linear function $f(x(\tau))$ in Eq.~(1) for 
the values  $p_1=0.8,p_2=1.33$.}
\label{w1}
\end{figure}

\begin{figure}[!ht]
\caption{Curves of Eq.~(10a) and Eq.~(10b) in the range of $-p_1 < x \leq p_1$.
The solid curve represents $\tau_1$ for $n=0,+1,+2$ and broken curve
represents  $\tau_2$ for $n=+1,+2$. The region enclosed between the line
$\tau=0$ and the curve $\tau=\tau_1(0,1.5b)$  is the only stable island (shaded
region).}
\label{fig1a}
\end{figure}

\begin{figure}[!ht]
\caption{Pseudochaotic attractor for $\tau=5.0, a=1.0, b=1.2, c=0.001, p_1=0.8, p_2=1.33$, 
$x(t)=0.9$, $t\in(-5,0)$. } 
\label{fig2}
\end{figure}

\begin{figure}[!ht]
\caption{Bifurcation diagrams of single cell DCNN for the parameters values
$a=0.16,b=0.2$  and $\tau=25.0$ when $p_2=1.33$ (a) for transients of the order
$1.0-2.0 \times 10^4$ and  (b) for transients now of the order $1.0\times
10^5$.}
\label{fig2a}
\end{figure}

\begin{figure}[!ht]
\caption{Bifurcation diagrams of single cell DCNN for the parameter values
$a=0.16,b=0.2$  and $\tau=25.0$ when $p_2=1.66$ (a) for transients of the order
$1.0\times 10^5$ and  (b) for transients of order $1.4\times10^6$.}
\label{fig3}
\end{figure}

\begin{figure}[!ht]
\caption{Bifurcation diagrams of single cell DCNN for the parameter values
$a=1.0,b=1.2,c=0.001 ,p_2=1.33$ and $\tau \in(2,29)$ (a) for transients of the
order $1.0\times 10^5$ and (b) for transients of the order $2.5\times 10^5$.}
\label{lypbif}
\end{figure}

\begin{figure}[!ht]
\caption{Global bifurcation diagrams of single cell DCNN for $\tau \in(0.5,30)$
and  $c \in(-0.16,0.16)$. The following colour codes are used to represent 
various regions, period-1 region -red, period-2 region -green, 3-blue, 
4-yellow, 5-magenta, 6-cyan, 7-gray, 8-copper, chaos-black and the  fixed
points-white. (a) $p_2=1.0$, (b) $p_2=1.33$ and (c) $p_2=1.66$.}
\label{fig4}
\end{figure}

\begin{figure}[!ht]
\caption{The first ten maximal Lyapunov exponents for the values
$a=1.0,b=1.2,c=0.001,p_2=1.33$  and $\tau \in(2,29)$.}
\label{fig1b}
\end{figure}

\begin{figure}[!ht]
\caption{Curves of Eq.~(\ref{eqfoura}) and Eq.~(\ref{eqfourb}) in the range of
$-p_1 < x \leq p_1$.  The solid curve represents $\tau_1$ for $n=0,+1,+2$ and
broken curve represents  $\tau_2$ for $n=+1,+2$. The region enclosed between
the line $\tau=0$ and the curve $\tau=\tau_1(0,1.5b)$  is the only stable
island (shaded region).}
\label{figad}
\end{figure}

\begin{figure}[!ht]
\caption{Bifurcation diagrams of coupled cell  defined by Eq.~(16). (a) For the
parameters values $a=0.16,b=0.2, d=0.2,e=0.2$ and $\tau=25.0$ for $p_2=1.33$
and for transients of the order $2.0 \times 10^5$ and  (b) For the same
parameter values as in (a) except for the transients of the order $1.75\times
10^6$.}
\label{figccwdb1}
\end{figure}

\begin{figure}[!ht]
\caption{Bifurcation diagrams of the two cell DCNN (a) For the parameters
values $a=0.16,b=0.2, d=0.2,e=0.2$ and $\tau=25.0$ when $p_2=1.66$  for
transients of the order $2.0 \times 10^5$ and  (b) For the same parameter
values as in (a) except for transients of the order $2.0\times 10^6$.}
\label{figccwdb2}
\end{figure}

\begin{figure}[!ht]
\caption{Intermittent behavior at the parameter values
$a=0.16,b=0.2,d=0.2,e=0.2$ and $\tau=25.0$  when $p_2=1.66$ for critical value
of the parameter $c_{crit}=0.345969$.}
\label{figint}
\end{figure}

\begin{figure}[!ht]
\caption{Mean laminar length $<l>$ versus $f=c_{crit}-c$.}
\label{figpl}
\end{figure}

\begin{figure}
\caption{Global bifurcation diagrams of coupled cell DCNN for $\tau \in
(0.1,30)$ and $c \in (0.3,1.4)$. The colour codes are the same as in Fig.~7.
(a) $p_2=1.0$, (b) $p_2=1.33$ and (c) $p_2=1.66$.}
\label{fig6}
\end{figure}

\end{document}